\documentclass[%
reprint,
superscriptaddress,
showpacs,preprintnumbers,
longbibliography,
 amsmath,amssymb,
 aps,
pra,
floatfix,
]{revtex4-1}

\usepackage{graphicx}% Include figure files
\usepackage{dcolumn}% Align table columns on decimal point
\usepackage{bm}% bold math
\usepackage{epsfig}
\usepackage{color}
\usepackage{textcomp}

\begin{document}
\title{On-chip Optical Squeezing}

\author{Avik Dutt}
\affiliation{School of Electrical and Computer Engineering, Cornell University, Ithaca, NY 14853, USA}

\author{Kevin Luke}
\affiliation{School of Electrical and Computer Engineering, Cornell University, Ithaca, NY 14853, USA}

\author{Sasikanth Manipatruni}
\affiliation{Exploratory Integrated Circuits, Intel Components Research, Intel Corp, Hillsboro, OR 97124, USA}

\author{Alexander L. Gaeta}
\affiliation{School of Applied and Engineering Physics, Cornell University, Ithaca, NY 14853, USA}
\affiliation{Kavli Institute at Cornell for Nanoscale Science, Cornell University, Ithaca, NY 14853, USA}

\author{Paulo Nussenzveig}
\affiliation{School of Electrical and Computer Engineering, Cornell University, Ithaca, NY 14853, USA}
\affiliation{Instituto de F\'isica, Universidade de S\~ao Paulo, P.O. Box 66318, 05315-970 S\~ao Paulo, Brazil}

\author{Michal Lipson}
\affiliation{School of Electrical and Computer Engineering, Cornell University, Ithaca, NY 14853, USA}
\affiliation{Kavli Institute at Cornell for Nanoscale Science, Cornell University, Ithaca, NY 14853, USA}
\email{ml292@cornell.edu}


\begin{abstract}
We present the first demonstration of 
 all-optical squeezing in an on-chip monolithically integrated CMOS-compatible platform. Our device consists of a low loss silicon nitride microring optical parametric oscillator (OPO) with a gigahertz cavity linewidth. We measure 1.7 dB (5 dB corrected for losses) of sub-shot noise quantum correlations between bright twin beams generated in the microring four-wave-mixing OPO pumped above threshold.
This experiment demonstrates a compact, robust, and scalable platform for quantum optics and quantum information experiments on-chip.
\end{abstract}

\maketitle

\section{Introduction}

Quantum properties of light can be used in a myriad of applications, ranging from enhanced 
sensing~\cite{goda_quantum-enhanced_2008, collaboration_gravitational_2011}, to spectroscopy~\cite{polzik_spectroscopy_1992, ribeiro_sub-shot-noise_1997}, 
to metrology~\cite{giovannetti_quantum-enhanced_2004, marino_absolute_2011}, and quantum information processing~\cite{braunstein_quantum_2005, obrien_photonic_2009}. Especially in 
the latter case, it would be desirable to have nonclassical light sources on compact platforms with very small footprint, high degree of 
confinement, low power operation, and compatibility with complementary metal oxide semiconductor (CMOS) technology. In addition to the possibility of leveraging on a mature fabrication technology that is already in place, such a CMOS compatible
platform would enable the integration of microelectronics with quantum photonics on the same substrate. Many efforts have been directed at the development of on-chip single-photon 
light sources~\cite{clemmen_continuous_2009, sharping_generation_2006, suhara_generation_2009, horn_monolithic_2012, davanco_telecommunications-band_2012, takesue2008generation}, detectors \cite{najafi_-chip_2015}, and logic gates~\cite{obrien_photonic_2009}. The development of brighter nonclassical light sources, such as squeezed light sources \cite{lvovsky_squeezed_2014}, on CMOS-compatible platforms has been lagging behind. 

Generation of squeezed states of light requires an optical nonlinearity and was initially demonstrated in several off-chip platforms including optical parametric oscillators (OPOs) using parametric down conversion~\cite{heidmann_observation_1987}, and in atomic vapors~\cite{slusher_observation_1985} and optical fibers~\cite{shelby_broad-band_1986} using four-wave mixing. In such nonlinear parametric processes two ``twin" beams are generated, called signal and idler beams, with strong\texttt{} quantum correlations in the intensities of the two beams, 
leading to noise in their intensity difference reduced below the standard quantum limit. In order to boost the effective optical nonlinearity it is customary to use a cavity. This introduces 
a tradeoff between high bandwidth and high squeezing. On the one hand, the cavity should have a high finesse, in order to achieve stronger squeezing. On the other hand, typically 
a high finesse implies low bandwidth, as is the case in free space OPOs. But for applications such as entanglement based quantum key distribution (QKD) in the continuous variable regime \cite{silberhorn_quantum_2002}, a large squeezing bandwidth is essential to ensure high data rates. Bright twin beam intensity difference squeezing has been recently demonstrated in whispering gallery mode crystalline resonators made of lithium niobate~\cite{furst_quantum_2011}, but the platform is not integrated on chip, and has a cavity bandwidth of 30 MHz. Recently, Ast \emph{et al.}~\cite{ast_high-bandwidth_2013} observed broadband squeezing over a GHz by using a low finesse cavity for field enhancement. However, as Ast \emph{et al.} point out~\cite{ast_continuous-wave_2012, ast_high-bandwidth_2013}, the consequences are a very high parametric oscillation threshold. Another approach to achieve 
strong squeezing is to use a pulsed pump, in order to achieve very high intensities and thus strong nonlinear behavior. Using a periodically poled lithium niobate waveguide, 
Eto \emph{et al.}~\cite{eto_efficient_2011} measured squeezing of -5~dB. In their work, since there is no cavity, there is no fundamental bandwidth tradeoff, the data rate limit being related to the repetition rate of the pulsed pump. However, CMOS-compatibility, and a high degree of confinement, which are desirable in making compact integrated photonic structures, is lacking. 
It should be noted that Safavi-Naeini \emph{et al.}~\cite{safavi-naeini_squeezed_2013} have reported 0.2 dB squeezing of light using a silicon optomechanical resonator, which, although CMOS compatible, is intrinsically restricted to a bandwidth of a few MHz around the mechanical resonance, and uses suspended structures operating at low temperatures, making it less robust compared to sources based on all-optical nonlinearities.

\begin{figure}
\includegraphics[width = .50\textwidth]{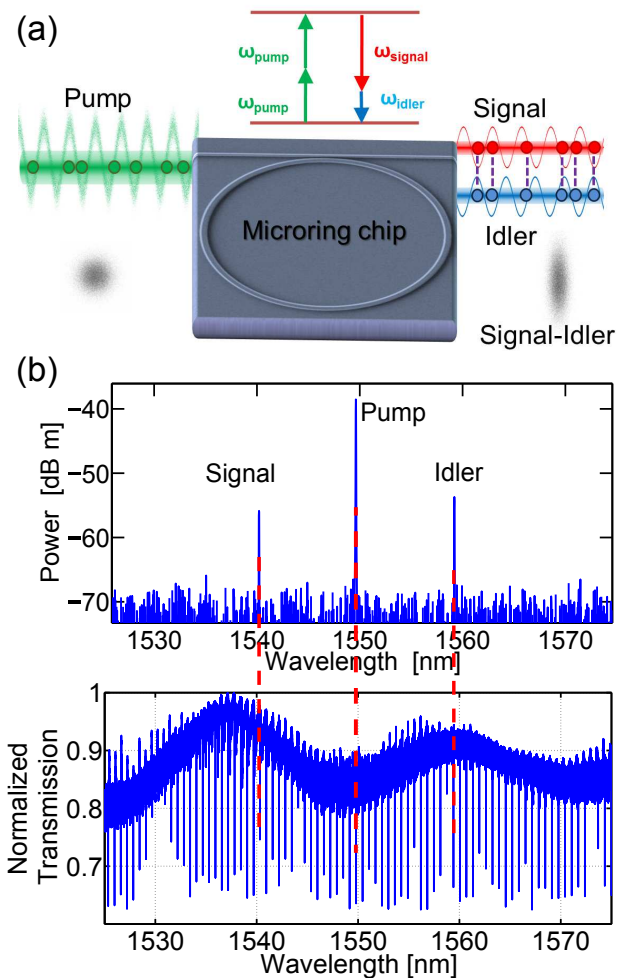}
\caption{(a) Schematic of the generation of intensity correlated signal and idler beams via four wave mixing in the on-chip microring optical parametric oscillator (OPO). (b) Top: Optical spectrum analyzer scan of the light coupled out of the chip, at a pump power just above the parametric oscillation threshold. The Si$_3$N$_4$ ring is pumped at 1549.6 nm, generating signal and idler modes at 1540.2 nm and 1559.2 nm, respectively, which are 15 cavity modes away from the pump wavelength, as can be seen in the bottom figure. Bottom: Transmission spectrum of the ring for transverse electric (TE) polarization.}
\label{fig:FWMprocess}
\end{figure}

In this article, we report the 
observation of all-optical squeezing in an on-chip monolithically integrated CMOS-compatible platform, generated in a micron-size silicon nitride oscillator~\cite{levy_cmos-compatible_2010} with gigahertz cavity linewidth. Owing to the small size of on-chip cavities, it is possible to obtain a large finesse and still have relatively wide cavity bandwidths. Large intra-cavity pump enhancement and strong squeezing can thus be obtained. Our OPO is based on the third order nonlinear process of four-wave mixing (FWM), as shown in Fig. \ref{fig:FWMprocess}.
% Consequently, this is also the first (to the best of our knowledge) generation of bright squeezed light using a singly pumped FWM system
 %(See \footnote{FWM-based bright squeezed light sources typically use a pump and a probe beam interacting in atomic vapors or a pulsed pump interacting with a seed signal in optical fibers. An unseeded source of bright squeezed light using FWM has not been demonstrated previously}). 
These devices consist of microring resonators fabricated on deposited films of Si$_3$N$_4$ (inset of Fig. \ref{fig:exptsetup}). Note that the OPO can generate in principle a very large number of beams spanning more than an octave~\cite{okawachi_octave-spanning_2011, foster_silicon-based_2011}. Here we generate squeezing by using a pump power that is just above the threshold, when only two modes oscillate.

\section{Methodology}
\subsection{Design}

We fabricate the microring resonators out of silicon nitride because of its high nonlinear refractive index~\cite{levy_cmos-compatible_2010} (n$_2 = 2.5 \times 10^{-15}\,  {\rm cm}^2\,  {\rm W}^{-1}$, about 10 times that of silica) and very low propagation loss ($< 0.5$ dB cm$^{-1}$). It should be noted that the nonlinear refractive index of silicon is an order of magnitude higher than that of Si$_3$N$_4$, but nonlinear losses such as two photon absorption and free carrier absorption at 1550 nm prohibit parametric oscillation in silicon. The devices are compact, with a bus waveguide length of 1.5 mm and a ring circumference of 1.8 mm, corresponding to a free spectral range (FSR) of 80 GHz.
{
In order to obtain FWM gain and optical parametric oscillation, the dispersion is engineered to be anomalous at the pump wavelength to compensate for the Kerr nonlinear phase shift due to the pump \cite{turner_tailored_2006, levy_cmos-compatible_2010}. We design the ring waveguide cross section to be 820 nm high and 1700 nm wide, resulting in a slightly anomalous group velocity dispersion at the pump wavelength of 1549.6 nm. %, as shown in Fig. \ref{fig:dispersion}.
}

In order to obtain significant squeezing, the cavity output coupling losses must be a large fraction of the overall losses. Relatively large losses are also desired for a large cavity bandwidth. On the other hand, a low pump power threshold is desirable, since it avoids detrimental thermal effects and also minimizes the influence of technical noise from the pump laser. The squeezing factor depends on the ratio between the internal ring losses, given by the ring's intrinsic quality factor ($Q_i$), and the coupling coefficient between the bus waveguide and the ring resonator, determined by the loaded quality factor ($Q_L$). The smaller the ratio of these losses, the better the squeezing obtained. {The squeezing factor relative to the shot noise level can be quantified as~\cite{brambilla_nondegenerate_1991,  castelli_quantum_1994, matsko_optical_2005, chembo_quantum_2014},
\begin{equation}
S(\Omega) = 1 - \frac{\eta_c \eta_d}{1 + \Omega^2 \tau_c^2} 
\end{equation}
where $\Omega$ is the sideband frequency, $\tau_c$ is the cavity photon lifetime, $\eta_d$ is the detection efficiency and $\eta_c = 1 - Q_L/Q_i$ is the ratio of the coupling losses to the total losses.}

The simultaneous requirement of large coupling losses and low threshold is a challenge, since the threshold is inversely proportional to the product of $Q_i$ and $Q_L$. We can meet these requirements by designing a ring with a very large $Q_i$, larger than 2 million, and operating in the highly overcoupled regime. The high intrinsic Q, achieved using the recently demonstrated fabrication process of thick Si$_3$N$_4$ deposition (Ref. \cite{luke_overcoming_2013}), enables the generation of the beams with a low pump power threshold of around 90 mW despite operating in this overcoupled regime. The corresponding intracavity power is estimated to be 7 W, which is an order of magnitude lower than the intracavity pump oscillation threshold of 65 W in Ref. \cite{ast_high-bandwidth_2013}. The loaded Q (200,000) is designed to be much lower than the intrinsic Q ($Q_i \approx 2$ million), facilitating a broad cavity linewidth of the order of 1 GHz.

The key to the realization of on-chip optical squeezing is the engineering of the dispersion and quality factors of the longitudinal modes of the microring resonator, to generate the twin beams at two well distinguished frequencies with low threshold pump powers. In contrast to many tabletop squeezing experiments, where the twin beams can usually be separated spatially based on their different polarizations, in our platform the beams are co-propagating in the same output waveguide with equal polarizations, and therefore in order to be distinguishable, the beams are required to have well separated frequencies. A delicate interplay between the group velocity dispersion and the quality factor of the ring, which together shape the FWM gain curve around the pump determines at which frequencies the twin beams will be generated \cite{lamont_route_2013, herr_universal_2012, chembo_quantum_2014}. {Specifically, the frequency difference between the pump and the first pair of oscillating signal and idler modes scales inversely as the square root of the second order dispersion ($D_2$) and the loaded quality factor($Q_L$): $\omega _{\rm pump} - \omega_{\rm signal} \propto \mathrm{FSR}/\sqrt{Q_L D_2}$}.  Here we engineer the structure to ensure that the signal and the idler beams are generated at two well distinguished wavelengths separated by around 20 nm, at 1540.2 and 1559.2 nm when the OPO is pumped at 1549.6 nm, enabling the beams to be spatially separated using a diffraction grating with very low loss. These two modes are 15 cavity resonances away from the pump wavelength, as can be seen from the transmission spectrum of the ring shown in Fig. \ref{fig:FWMprocess}(b). The large wavelength separation between twin beams not only enables spatial separation of the twin beams, but  also enables filtering of the pump wavelength, which is essential to mitigate the effects of the residual pump from affecting the twin beam squeezing measurements.

\subsection{Device Fabrication}
{
The Si$_3$N$_4$ waveguides and ring resonators were fabricated in 820 nm-thick silicon nitride films to provide low loss and high optical confinement. A 4 $\rm \mu$m thick silicon dioxide layer was first thermally grown on a virgin silicon wafer as an under cladding. The nitride layer was grown using low pressure chemical vapor deposition (LPCVD) in two steps of 400 nm each, interleaved with annealing in a nitrogen atmosphere for 3 hours at 1200 \textdegree C. The devices were patterned with electron beam lithography using MaN-2403 resist. After exposure and development, the resist was post-exposure baked for 5 min. at 115 \textdegree C, and etched in an inductively coupled plasma reactive ion etcher (ICP RIE) using CHF$_3$/O$_2$ chemistry. The devices were finally clad with 250 nm of high temperature oxide (HTO) deposited at 800 \textdegree C, followed by 2  $\mu$m of silicon dioxide using plasma enhanced chemical vapor deposition. Further details about the fabrication process can be found in \cite{luke_overcoming_2013}.
}

\begin{figure}
\includegraphics[width = .50\textwidth]{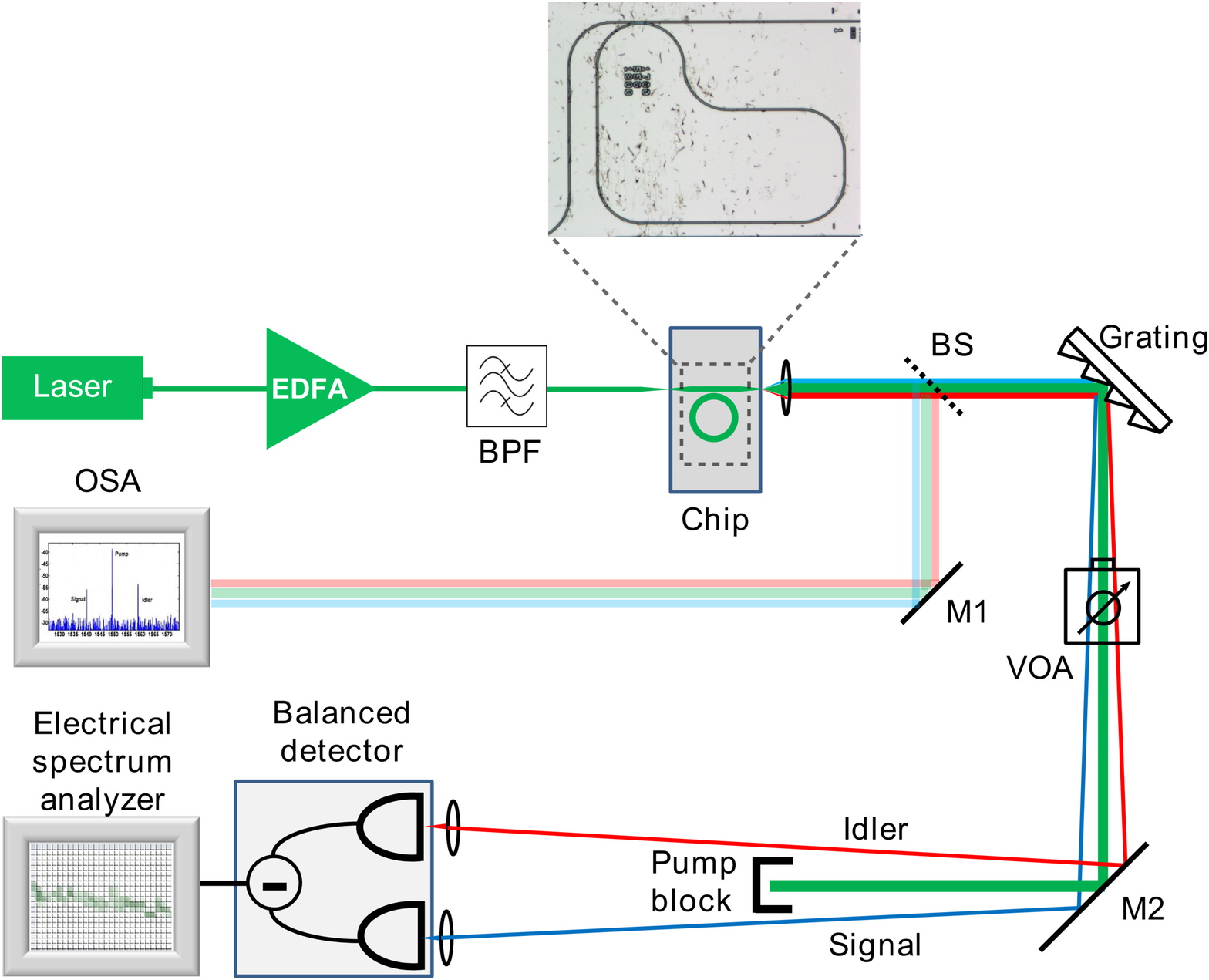}
\caption{Experimental setup. The on-chip microring resonator is pumped by a continuous wave laser followed by an erbium doped fiber amplifier (EDFA) from JDS Uniphase and a bandpass filter (BPF). To ensure efficient coupling in and out of the chip, inverse tapers are used \cite{almeida_nanotaper_2003}. A lensed fiber is used for input coupling. The output from the waveguide is collected with a high NA objective lens. The diffraction grating spatially separates the pump and the twin beams. After blocking the pump, the beams are focused on a balanced detector. The RF output of the balanced detector is sent to an electrical spectrum analyzer. A small fraction of the beams is sent to an optical spectrum analyzer (OSA) to monitor the onset of parametric oscillation. The inset shows a microscope image of the ring resonator coupled to the bus waveguide. The shape of the fabricated ring is non-circular to fit the resonator within one field of the electron beam lithography tool and hence avoid stitching errors.
EDFA: Erbium doped fiber amplifier. BPF: Bandpass filter. OSA: Optical spectrum analyzer. VOA: Variable optical attenuator. M1, M2: mirrors. BS: Beam splitter}
\label{fig:exptsetup}
\end{figure}

\subsection{Experimental setup}
 The experimental setup used to measure the squeezing is shown in Fig. \ref{fig:exptsetup}. The ring was pumped by a continuous wave, tunable laser (New Focus Velocity 6328) followed by an erbium doped fiber amplifier (JDSU OAB1552+20FA6) and a bandpass filter (BPF) with a bandwidth of 9 nm to reduce the amplified spontaneous emission noise generated by the EDFA.  Light is evanescently coupled to the microring through a waveguide. The output from the waveguide was collected with a high NA (NA = 0.25) objective lens, leading to a loss of less than 1 dB. A diffraction grating with an efficiency of 85\% was used to spatially separate the pump and the two beams. After blocking the pump, the beams were focused on the two inputs of a Thorlabs PDB 150C balanced detector. The balanced detection system consists of a pair of well-matched InGaAs photodiodes with a quantum efficiency of 80\%, followed by a low noise transimpedance amplifier to amplify the difference in photocurrents between the detectors. To observe the variance of the intensity difference noise between the two balanced detectors, the RF output of the balanced detection system was sent to an electronic spectrum analyzer. The detection system had a bandwidth of 5 MHz and a transimpedance gain of 10$^5$ V/A. A small fraction (5\%) of the light coming out of the chip was sent to an optical spectrum analyzer (OSA) to observe the onset of parametric oscillation. The setup includes a variable optical attenuator (VOA) to calibrate the shot noise level and to check the degradation of squeezing with attenuation of the twin beams.

\section{Results and discussion}

We measure the squeezing in intensity difference of the two beams generated when the triply resonant OPO is pumped above threshold by comparing the noise level to the shot noise level generated from the coherent pump source.

\subsection{Shot noise calibration}

\begin{figure} %[!hb]
\includegraphics[width = .5\textwidth]{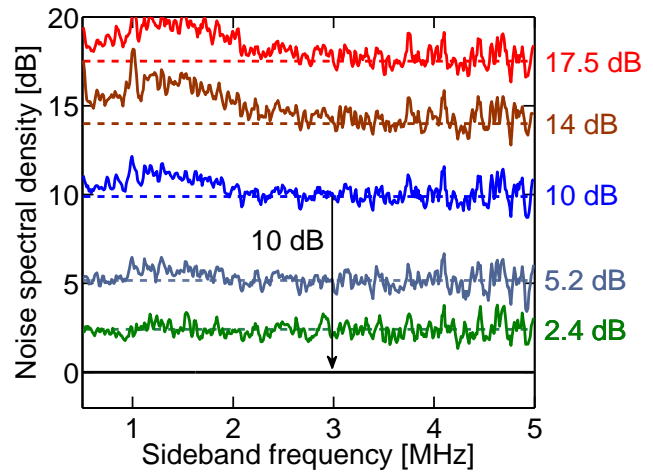}
\caption{Linearity of the balanced detection system from 0.5 MHz to 5 MHz. The data have been dark noise corrected and normalized to  the shot noise spectrum at a base local oscillator (LO) power of $P_0 = 7.8\, \mu $W, corresponding to the horizontal solid line at 0 dB. The dashed lines depict the mean optical power in the corresponding spectrum referenced to the base LO power $P_0$, in dB. There is significant technical noise at low frequencies, which is clearly evident at relative powers above 10 dB. The data have more fluctuations at high frequencies due to higher electronic noise and consequently lower dark noise clearance for the base LO power $P_0$. The dark noise varies from 5 dB below the 0 dB reference at 0.5 MHz to 1 dB below the reference at 5 MHz. Around 3 MHz, the detection system is linear over more than 20 dB, which is also substantiated by the linearity of the shot noise calibration measurement in Fig.~\ref{fig:shotnoise}.
}
\label{fig:spectrashot}
\end{figure}

\begin{figure}
\includegraphics[width = .50\textwidth]{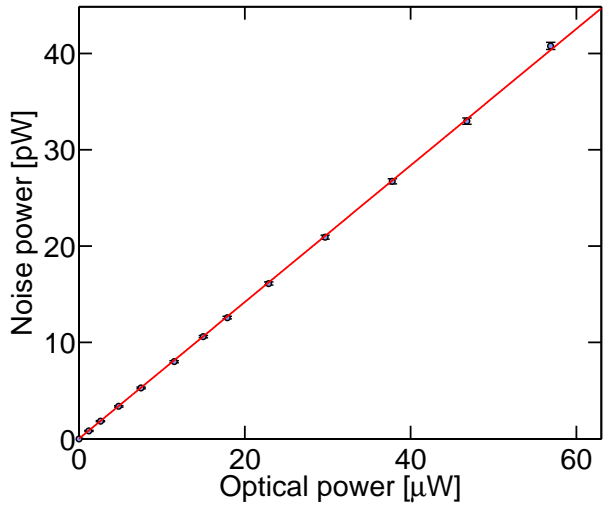}
\caption{Shot noise calibration, showing the linearity of the noise power in the photocurrent difference with optical power, as expected from theory. The data are taken at a Fourier sideband frequency of 3 MHz, using a spectrum analyzer with a resolution bandwidth of 30 kHz and a video bandwidth of 100 Hz. Data points have been corrected by subtracting the electronic noise of the balanced detection system. Error bars are extracted from the raw data.
}
\label{fig:shotnoise}
\end{figure}

{We calibrate the shot noise level by splitting the pump beam on a 50:50 beam splitter, directing the two halves to the two detectors and monitoring the dc and ac components of the balanced detector output simultaneously.  The common mode rejection ratio (CMRR) of the detectors was independently measured to be more than 27 dB. The laser is detuned far from the microring cavity resonance during shot noise calibration measurements. The electronic noise of the detection system is measured by blocking all light incident on the detectors. As seen from Fig. \ref{fig:spectrashot}, the detection system is linear over 20 dB near a sideband frequency of 3 MHz. There is excess technical noise at frequencies below 2 MHz, which cannot be rejected completely by the finite CMRR of the detection system. Focusing on the noise at 3 MHz, it can be seen from Fig. \ref{fig:shotnoise} that the shot noise level is linearly proportional to the optical power in each beam, in accordance with theory.
}

\subsection{Squeezing measurements}

\begin{figure}
\includegraphics[width = .50\textwidth]{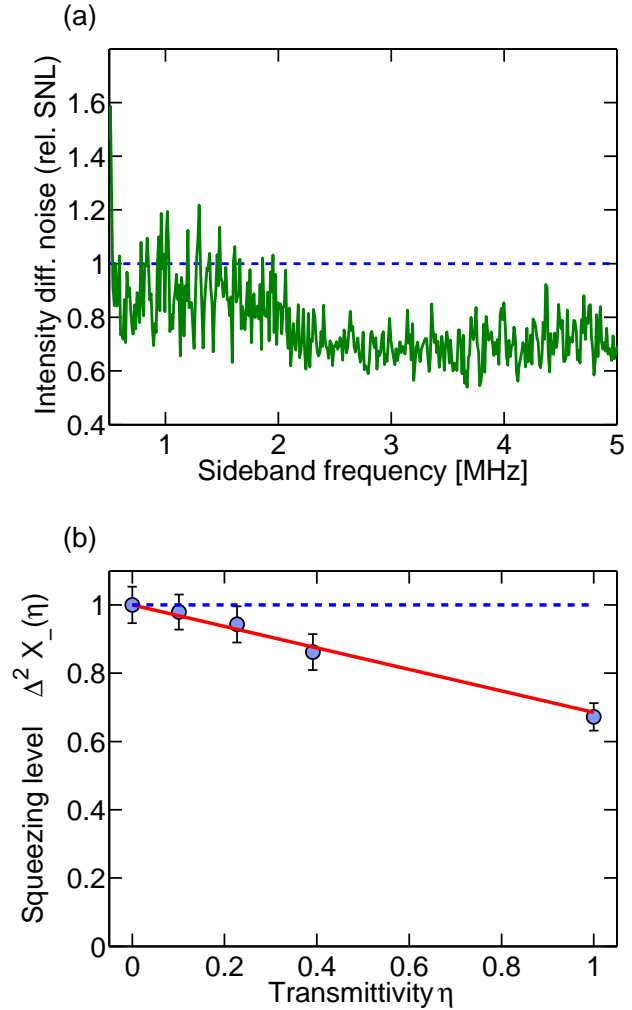}
\caption{Intensity difference squeezing. a) The variance of the signal and idler photocurrent difference, normalized to the shot noise level (SNL), on a linear scale. The dashed line at 1 represents the SNL, and the solid line represents the intensity difference fluctuations. The dark noise clearance was 16 dB at 0.5 MHz, and decreased to 5 dB at 5 MHz. The data was taken with a pump power of 93 mW. The power in the signal and idler beams is 45 $\rm \mu$W, which requires low dark noise detectors, commercially unavailable at higher frequencies. {Squeezing at low frequencies is masked by excess technical noise below $\sim$ 2 MHz, that can be clearly seen in Fig.~\ref{fig:spectrashot}}. b) Variation of squeezing with attenuation at a sideband frequency of 3 MHz. The x axis is the transmittivity of the variable optical attenuator, $\eta$. The y-axis is the intensity difference noise in shot noise units, that is, the variance of the signal and idler photocurrent difference compared to the shot noise level. Error bars are determined from the standard deviation of the measured data points.
}
\label{fig:squeezing}
\end{figure}

 We observe more than 30 \% sub-shot noise quantum intensity correlations between the twin beams generated when the pump laser is tuned to an on-chip resonance at 1549.6 nm. Twin beam intensity difference squeezing measurements are presented in Fig. \ref{fig:squeezing}(a). The  solid line corresponds to the signal-idler intensity correlation measurement, which is below the shot noise level, demonstrating clear intensity difference squeezing. These measurements were taken at Fourier sideband frequencies from 0.5 to 5 MHz, using a spectrum analyzer with a resolution bandwidth of 30 kHz and a video bandwidth of 100 Hz. Squeezing is not observed at very low frequencies, owing to technical noise in the pump laser. The on-chip OPOs used here could in principle act as a platform for generating large squeezing factors over broad bandwidths due to the highly overcoupled design of the rings. 
 
 {We confirm that the squeezing factor degrades linearly with increasing attenuation, and the intensity difference noise approaches the shot noise level for high attenuation, as is typical of squeezed states (Fig.~\ref{fig:squeezing}(b)). 
  %The intensity difference squeezing factor is expected to decrease linearly  as a function of decreasing transmission, as is typical of squeezed states, reaching the shot noise level for very low transmission values.
By using the variable optical attenuator (VOA) in the experimental setup (Fig.~\ref{fig:exptsetup}), we measure the influence of decreasing transmission through the VOA on the squeezed twin beams. The  dependence of the variance of the intensity difference noise, normalized to the corresponding shot noise level, can be modeled by mixing the unattenuated two-mode squeezed state with a vacuum state, on a beam splitter with transmittivity $\eta$:
 
 \begin{equation}
 \Delta^2 X_-(\eta) = \eta\,  \Delta^2 X^{(0)}_- + (1-\eta)\,  \Delta^2 X_v
 \end{equation}
 where the left hand side represents the variance after attenuation, $\Delta^2 X_-^{(0)}$ represents the unattenuated variance (i.e. before the VOA), and $X_v = 1$ is the variance of vacuum. All variances are normalized to the shot noise level at the corresponding power.}
 
  The observed noise reduction of 1.7 $\pm$ 0.4 dB corresponds to a generated squeezing of 5 dB when corrected for detection losses and the non-ideal quantum efficiency of the detectors. This is less than the 10 dB of squeezing expected from the ratio of $Q_i$ and $Q_L$ owing to residual excess pump noise of 25 dB relative to the shot noise level, and the possible rotation of the optimally squeezed quadratures by the process of FWM \cite{mckinstrie_schmidt_2013, corzo_rotation_2013}. Furthermore, higher intrinsic quality factors of 7 million have been demonstrated in silicon nitride rings~\cite{luke_overcoming_2013}, resulting in a lower oscillation threshold, which not only  helps in reducing excess noise in the pump, but also leads to a higher ratio of $Q_i$ to $Q_L$. It should thus be possible to reach much stronger noise reductions in this platform. Fundamentally the on-chip OPOs are expected to exhibit squeezing over GHz bandwidths in view of the broad linewidth of the cavity. We have demonstrated here squeezing in the MHz range, limited only by the bandwidth of our low dark noise detectors. 
%It should be noted that commercially available detectors at 1550 nm have insufficient dark noise clearance at GHz sideband frequencies at optical powers of a few tens of $\mu$W, corresponding to the power in the generated twin beams just above threshold.

\section{Conclusions and outlook}

These results constitute an experimental demonstration of all-optical squeezing in an integrated CMOS compatible platform.
Our source generates bright squeezed light using a singly pumped FWM process, in contrast to other sources of above threshold squeezing which utilize parametric down-conversion. Since FWM is based on the third order nonlinearity, the technique presented here can be extended to several different material platforms, in contrast to the more restrictive second order nonlinearity only found in non-centrosymmetric materials. For example, FWM oscillation has been reported in silica~\cite{delhaye_optical_2007}, crystalline fluorides~\cite{savchenkov_tunable_2008}, hydex~\cite{razzari_cmos-compatible_2010}, aluminum nitride~\cite{jung_optical_2013} and diamond~\cite{hausmann_diamond_2014}. Our demonstration paves the way for a myriad of on-chip quantum optics experiments over broad bandwidths in a scalable, compact and robust platform. An experiment to measure phase anticorrelations of the twin beams is under way, enabling a demonstration of continuous-variable quantum entanglement~\cite{villar_generation_2005}. This will open the way to realize deterministic quantum information protocols at very high speeds. In future, the frequency separation of the twin beams can be done on chip using ring-resonator-based add-drop filters tuned to the signal, idler and pump wavelengths so that they are demultiplexed to different waveguides~\cite{barwicz_microring-resonator-based_2004}. This opens up the possibility to cascade the on-chip OPO with photonic structures to manipulate squeezed and entangled states of light generated by the OPO, further emphasizing the highly scalable nature of our platform.
The introduction of such non-classical light sources into future data communications by leveraging the mature infrastructure of microelectronics, currently being introduced into silicon photonics~\cite{jalali_silicon_2006}, is a very promising avenue to be explored.

\begin{acknowledgments}

We acknowledge fruitful discussions with Vivek Venkataraman, Alessandro S. Villar, Jaime Cardenas, Carl Poitras, Yoshitomo Okawachi, K. Saha, St\'{e}phane Clemmen and Marcelo Martinelli. The authors gratefully acknowledge support from DARPA for award \#W911NF-11-1-0202 supervised by Dr. Jamil Abo-Shaeer, and from AFOSR for award \#BAA-AFOSR-2012-02 supervised by Dr. Enrique Parra. P.N. acknowledges support from Funda\c{c}\~{a}o de Amparo \`{a} Pesquisa do Estado de S\~ao Paulo (FAPESP grant \#2011/12140-6). This work was performed in part at the Cornell NanoScale Facility, a member of the National Nanotechnology Infrastructure Network, which is supported by the National Science Foundation (Grant ECCS-0335765). This work made use of the Cornell Center for Materials Research Shared Facilities which are supported through the NSF MRSEC program (DMR-1120296).

\end{acknowledgments}

%\bibliography{MyLibrary_2015_03_19}

%\end{document}

%\begin{thebibliography}{99}


\end{document}